# Virtual Reality in Social Media: A New Era of Immersive Social Interactions

Priyanshu Chaubey
U.G. Student, Department of Computer Science Engineering, Parul University, Vadodara, Gujarat, India City
solvewitcode@gmail.com

## Abstract:

In order to explore the transformative potential of immersive experiences for user engagement, interaction, and content production, this paper examines the integration of Virtual Reality (VR) capabilities in social media apps. Examining sites like Facebook Horizon, VRChat, and the use of AR/VR in social apps like Snapchat, the study examines the current state of VR in social media. It also discusses the obstacles to VR acceptance, like hardware constraints and accessibility, and forecasts future developments in this field. This paper's conclusion assesses the advantages and challenges of integrating social media with virtual reality, offering insights into how these technologies will influence digital communication.

Human communication has been profoundly changed by social media, which allows users to engage in previously unheard-of ways, such as text-based conversations, video chats, and live streaming. The digital landscape has started to change in recent years as a result of the introduction of Virtual Reality (VR) to these platforms. Instead of using conventional 2D screens, VR offers a completely immersive experience that lets users interact with content and one another in 3D spaces. This study examines the integration of virtual reality (VR) technology into social media applications, evaluating their potential to provide more dynamic and captivating digital spaces. Globally, social media sites like Facebook, Instagram, and Twitter have already changed the nature of communication. Immersion technologies like virtual reality (VR) represent the next stage, though, as they have the ability to change how we interact, connect, and share in social settings in addition to improving user experience.

## Literature Review:

The evolution of social media started with straightforward text-based sites like MySpace and progressed to sophisticated multimedia experiences with Facebook, Instagram, and TikTok. These platforms concentrated on giving users the means to connect, communicate, and exchange content. As live-streaming and augmented reality (AR) filters gained popularity, social media became more dynamic and engaging. The next big thing that could make these virtual worlds come to life is the integration of virtual reality (VR) into social media applications. Virtual Reality in Social Media: Although VR has traditionally been linked to entertainment and games, it is also becoming more and more prevalent in social media. Platforms like AltspaceVR, VRChat, and Facebook Horizon (formerly Meta Horizon) offer completely immersive virtual worlds where users may communicate with one another in three dimensions. According to research, virtual reality (VR) can produce a sensation of "presence," or the perception of being physically present in a virtual environment, which has the potential to greatly improve online social interactions. However, there are still issues with VR's incorporation into social media, such as the requirement for specialist technology, user adjustment, and entrance costs. Research Questions: The following important research questions are the focus of this study: What effects does the incorporation of virtual reality features into social media platforms have on user happiness and engagement? What accessibility and technological obstacles prevent VR from being widely used on social media? In terms of community development, content sharing, and user interaction, how are VR social networks different from conventional social media? What possible ethical and privacy issues are raised by the use of virtual reality on social media?

## Methodology:

**Research and Analysis**

Incorporating **ExoPlayer** into a social media app can offer users a unique and immersive experience. Our research outlines the development process for the social media app with integrated video playback features, utilizing **Android Studio** and **Firebase** for a comprehensive solution.

**Market Research:**

With the advent of 5G technology, the demand for high-quality video playback applications has surged. **ExoPlayer** is increasingly being recognized for its capabilities in providing seamless video streaming experiences across various platforms.

**Competitor Analysis:**

- **ExoPlayer:** As a robust media player, ExoPlayer enables smooth playback of a variety of media formats. Its adaptive streaming capabilities make it an ideal choice for applications that require dynamic video content delivery.

- **Other Video Streaming Platforms:** We examined popular platforms that utilize advanced media players for content delivery, focusing on their user experience and engagement strategies.

**Content Strategy and Creation:**

Users can create their own video content and share it on our app, utilizing **ExoPlayer** to ensure high-quality playback. By leveraging ExoPlayer's adaptive streaming, we aim to provide a reliable video viewing experience, regardless of network conditions.

## Requirement Gathering

During the requirement-gathering phase, we identified essential features for integrating **ExoPlayer**

into our application:

- Support for multiple video formats.
- Customizable playback controls.
- Offline Viewing capabilities.

## Design and Planning

In the design phase, we focused on how to implement **ExoPlayer** to manage video playback effectively within **Android Studio**. The architecture includes:

- A user interface component for video playback, allowing seamless integration with messaging features.

- Storage management for user-uploaded videos, ensuring smooth access and retrieval through

  **ExoPlayer**.

- Backend support through **Firebase**, facilitating secure storage, user authentication, and real- time data synchronization.

## Development

The application was developed using **Android Studio** as the primary IDE, with **ExoPlayer** for all video playback functionalities. The development process included:

- Implementing **ExoPlayer** for adaptive streaming and handling various media formats.
- Utilizing **Firebase** for user authentication and storing media files securely.
- Creating user-friendly interfaces in **Android Studio** for easy navigation and interaction with video content.
- Ensuring efficient media caching to enhance playback performance.

## Testing and Quality Assurance

Testing was conducted to ensure that **ExoPlayer** functions effectively within the application:

- Unit tests focused on video loading and playback functionalities.
- Integration tests verified the seamless interaction between **ExoPlayer**, **Firebase**, and other application components.
- Usability testing gathered user feedback on the video playback experience, making adjustments to enhance user satisfaction.

## Deployment and Launch

The application featuring **ExoPlayer** for video playback and **Firebase** for backend services was deployed to app stores, providing users access to high-quality video content seamlessly integrated into their social interactions.

### Monitoring and Maintenance

Post-launch, the application will continuously monitor **ExoPlayer** performance and **Firebase** data interactions, addressing any issues related to video playback and ensuring a consistently high-quality user experience.

## Case Studies:

On the following platforms, two case studies will be carried out:

Create avatars, explore virtual worlds, and engage with other people on Meta Horizon (previously Facebook Horizon), a social VR platform created by Meta (formerly Facebook). VRChat: An online community where users may connect, communicate, and engage in activities with modifiable avatars. Many people utilize the site to host virtual gatherings and socialize.

Comparing the VR settings to conventional social media platforms, both case studies will examine user interaction, engagement, and community creation.

## Analysis and Discussion:

The results of the surveys, case studies, and literature research will be examined in this part.

User Engagement and Social Interactions: The capacity to build immersive environments that foster a sense of community among users is one of the main advantages of incorporating virtual reality (VR) into social media. Users may communicate in real time with highly customisable avatars on platforms like VRChat, which creates a sense of community that is impossible to achieve on more conventional social media sites. Because VR is immersive and enables more organic social connections, users report higher levels of engagement in these situations. Users can participate in virtual hangouts, gaming, and event attendance, for instance, which simulates face-to-face interaction.

Technological Barriers: A number of technological obstacles stand in the way of VR's possible integration into social media. Specialized equipment, such VR headsets (like the Oculus Quest or HTC Vive), are necessary for high-quality VR experiences, but they can be costly. Furthermore, individuals with slower internet connections or lower-end devices may not be able to use VR technologies because to their high bandwidth and processing power requirements. Therefore, it can take some time before VR is widely used in social media.

Privacy and Ethical Issues: Because virtual reality is so immersive, privacy issues are brought up. In conventional social media, people post videos, images, and text that contain personal information. Users may, however, provide more specific details about their preferences, activities, and even physical motions in virtual reality settings. Consent and data security are issues because this data may be used for targeted advertising or other uses. Additionally, the anonymity that virtual reality avatars offer may raise moral questions about harassment and the dissemination of offensive material.

## Conclusion

In conclusion, the incorporation of virtual reality (VR) into social media platforms signifies a substantial change in the way people communicate and participate online. Although there is a lot of potential for developing dynamic, immersive social spaces, there are a number of obstacles that need to be overcome, such as privacy issues, accessibility issues, and technological constraints. Through the creation of more engaging and participatory experiences, VR-enhanced social media platforms such as Meta Horizon and VRChat have demonstrated the potential to revolutionize online communities. However, removing the financial and accessibility obstacles would be necessary for VR to become widely used. VR technology has the potential to completely transform online socialization, communication, and teamwork as it becomes more widely available and affordable.